\documentclass[sigconf]{acmart}

\usepackage{graphicx}
\usepackage{caption}
\usepackage{subcaption}
\usepackage{dirtytalk}

\AtBeginDocument{%
  \providecommand\BibTeX{{%
    \normalfont B\kern-0.5em{\scshape i\kern-0.25em b}\kern-0.8em\TeX}}}

\setcopyright{acmcopyright}
\copyrightyear{2020}
\acmYear{2020}

\acmConference[PervasiveHealth]{Pervasive Health}{2020}{Atlanta, GA}
\acmBooktitle{PervasiveHealth, 2020}
\acmPrice{15.00}
\acmISBN{978-1-4503-XXXX-X/20/05}

\begin{document}


\title{Understanding Reflection Needs for Personal Health Data in Diabetes}

\author{Temiloluwa Prioleau}
\affiliation{
  \institution{Dartmouth College}
  \streetaddress{9 Maynard Street}
  \city{Hanover}
  \state{NH}
  \postcode{03755}
}

\author{Ashutosh Sabharwal}
\affiliation{
  \institution{Rice University}
  \streetaddress{6100 Main St.}
  \city{Houston}
  \state{TX}
  \postcode{77005}
}

\author{Madhuri M. Vasudevan}
\affiliation{
  \institution{Baylor College of Medicine}
  \city{Houston}
  \state{TX}
  \postcode{77030}
}

\renewcommand{\shortauthors}{Prioleau et al.}

\begin{abstract}
To empower users of wearable medical devices, it is important to enable methods that facilitate reflection on previous care to improve future outcomes. In this work, we conducted a two-phase user-study involving patients, caregivers, and clinicians to understand gaps in current approaches that support reflection and user needs for new solutions. Our results show that users desire to have \textit{specific summarization metrics}, \textit{solutions that minimize cognitive effort}, and \textit{solutions that enable data integration} to support meaningful reflection on diabetes management. In addition, we developed and evaluated a visualization called PixelGrid that presents key metrics in a matrix-based plot. Majority of users (84\%) found the matrix-based approach to be useful for identifying salient patterns related to certain times and days in blood glucose data. Through our evaluation we identified that users desire data visualization solutions with \textit{complementary textual descriptors}, \textit{concise and flexible} presentation, \textit{contextually-fitting} content, and \textit{informative and actionable} insights. Directions for future research on tools that automate pattern discovery, detect abnormalities, and provide recommendations to improve care were also identified.

\end{abstract}




\maketitle

\section{Introduction}

\begin{figure}
	\includegraphics[width=0.6 \linewidth]{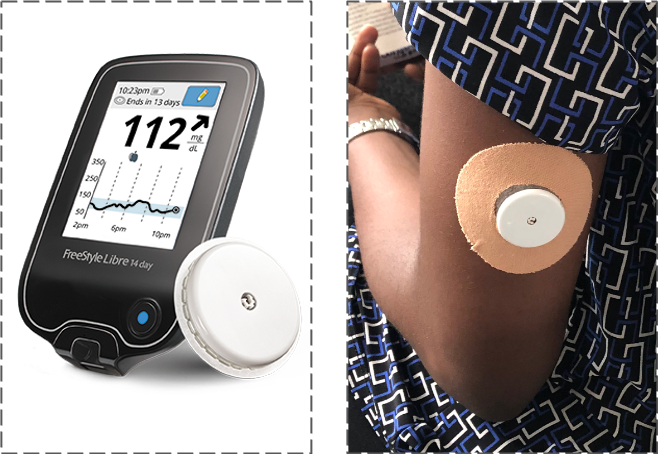}
	\caption{Example of a Continouous Glucose Monitor (FreeStyle Libre \cite{FreeStyleLibre}) - Wearable Medical Device.}
	\label{fig:FreeStyleLibre}
\end{figure}

Diabetes is the 7th leading cause of death in the United States \cite{CDC_Diabetes2017}. It is characterized by impaired glucose metabolism yielding frequent high and low blood glucose (BG) levels that increase the risk of long-term macro- and micro-vascular complications \cite{ADA2017}. In addition to prescribed medications and/or insulin use, management of diabetes also benefits from frequent monitoring of BG levels and a remedy action when blood glucose levels are above or below the healthy range. Persons on intensive insulin therapy, such as with Type 1 Diabetes (T1D), are often good candidates for continuous glucose monitors (CGM) to achieve 24/7 tracking of their BG levels. Figure \ref{fig:FreeStyleLibre} shows an example of a CGM worn on the back of a user's arm and the blood glucose value at a given time.

\par CGM use has been shown to improve diabetes management and reduce adverse events such as hypoglycemia also known as low blood glcuose \cite{Bergenstal2013}, \cite {Danne2017}, \cite{Fonseca2016}, \cite{vhaduri2020adherence}. However, a major barrier to optimal CGM use is in limitations of data delivery and reporting tools that enable effective interpretation \cite{Bergenstal2013}, \cite{Fonseca2016}. From the perspective of a patient or user, self-tracking using CGMs present similar challenges as other personal health devices including challenges with data interpretation and application to care \cite{Choe2014}, \cite{West2016}. Over the course of days and weeks, a patient's CGM dataset grows quickly and it becomes a challenge to identify salient information from the large dataset. The American Diabetes Association (ADA) recommends two to four medical visits on an annual basis for persons with diabetes \cite{ADA2017}. This translates to an adherent patient visiting with their healthcare provider every 3 to 6 months. From a physician's perspective, it is rather difficult to review, digest, and interpret 3 to 6 months worth of a patient's CGM data during short clinical visits (often about 20 minutes long). Therefore, physicians tend to review only the two most recent weeks of CGM data to assess glucose management since the last visit \cite{Carlson2017clinical}.


\par In this paper, we investigate user-needs with respect to decision-support tools that can facilitate reflection on personal health data from CGMs to guide and improve future diabetes management strategy.  In addition, we design and evaluate a visualization approach for CGM data to enable recognition of temporal patterns, and facilitate introspection on behavior over both short and long time periods. To achieve our objective, we first conducted a needs assessment survey to understand: 1) what users focus on when they review retrospective blood glucose data? and 2) what users view as existing gaps in state-of-the-art data presentation tools? From this study, we identified 3 important things that users of such personal health data (i.e. patients, caregivers, and clinicians) desire to support meaningful reflection on health, including \textit{specific summarization metrics, solutions that minimize cognitive effort, and solutions that enable data integration}. Specific to diabetes management, participants identified specific summarization metrics such as time-in-range, patterns/trends of highs and lows, average blood glucose, and trends during certain times/hours/days. Additionally, users identified gaps in state-of-art data reporting tools for diabetes care including solutions that automatically extract patterns, detect changes, and suggest recommendations to mitigate suboptimal outcomes, and solutions that integrate pertinent data from multiple sources (e.g. continuous glucose monitors and activity trackers).

Following this, we conducted a 2-phase iterative design and evaluation of matrix-based visualization scheme which we call PixelGrid. Input from a Phase 1 study was used to revise the design for further evaluation in Phase 2. A key result is that majority of the participants (84\% across both studies) found PixelGrid useful for identifying patterns as it relates to certain times/days. A total of 31 patterns were identified across all participants in Phase 2 testing. Some subjects also found PixelGrid useful in the process of pin-pointing habits that contribute to management gaps and informing positive modifications to treatment regimen. Participants also provided input to further develop decision-support tools for improved management of diabetes. Given that PixelGrid is a departure from standard visualization tools used in practice (i.e. daily overlay plots), both in layout and color scheme, it was expected that participants encountered some usability challenges considering that they had \emph{no prior training/exposure} to new visualization.

\par Through our design and evaluation process, we identified four important things that users desire in data visualization tools to support meaningful reflection on personal health data including, \textit{complementary textual descriptors}, \textit{concise and flexible} presentation, \textit{contextually-fitting} content, and \textit{informative and actionable} insights. While this paper focuses on diabetes as one application space, we envision that insights from this study can inform development of decision-support tools in other domains that benefit from personal health data. Recent literature shows that there is a surge in research and development of wearable health monitoring systems \cite{Chan2012}, \cite{Pantelopoulos2010}. For example, wearable electrocardiograms are becoming prevalent for continuous monitoring and treatment of arrhythmia (also known as irregular heartbeat) and several commercial products now exist such as iRhythm's Zio XT \cite{iRhythmZioXT}. As such wearable medical devices are becoming commonplace for health monitoring, hence, data delivery and reporting tools are critical to support reflection on previous data and improve future outcomes. 




\section{Related Work}


In this section, we present background and ongoing research on the topics of personal informatics, data visualization in health, and decision support in diabetes management.

\subsection{Personal Informatics}
Personal informatics is the process of building knowledge through self-logged data for reflection and discovery, to drive change in activities and behaviors, and to improve health \cite{choe2015characterizing}, \cite{Choe2014}, \cite{Li2011_ubicomp}, \cite{vhaduri2020adherence}, \cite{Wolf2009_KnowThyself}. In recent years, advancements in wearable and mobile systems are facilitating the transition from manual tracking to technology-driven monitoring. Sensors and systems have been developed for tracking physical activity, food intake, mental health, sleep, and much more \cite{ZhenyuChen2013}, \cite{Lara2013}, \cite{PrioleauTBME2017}, \cite{Wang_UbiComp2016}. The value of self-tracking and introspective is well established  \cite{Birkavs_QS2016}, \cite{Wolf2009_KnowThyself}. Toward designing effective technology that enables self-reflection, Li et al. \cite{Li2011_ubicomp} identified six things people want to understand from their personal data including: status, history, goals, discrepancies, context, and factors. The authors also introduced the phases of \textit{discovery} and \textit{maintenance} that a user often transitions to-and-fro, as they set and strive for goals. Other research supports the importance of designing persuasive technology to target various stages that individuals progress through in the decision-making process of intentionally modifying or changing a lifestyle behavior \cite{Consolvo2009}. This includes enabling an appropriate balance between technology and the human-in-the-loop \cite{Li_CHI2010}. 

\par For persons with ailments (e.g. chronic diseases), personal informatics can be particularly important to support the notion of \textit{patient-centered} care in a two-fold way. Firstly, personal informatics can help a patient better understand their health status on a continuous basis, support self-management, and motivate behavior change to improve outcomes. Mamykina et al. describe this process in a sensemaking framework that includes \say{perception of new information related to health and wellness, development of inferences that inform selection of actions, and carrying out daily activities in response to new information} \cite{Mamykina2015adopting}. A similar sensemaking framework to inform the design of diabetes decision support systems is also presented by Katz et al. \cite{katz2018designing}. Secondly, personal informatics can contribute to clinical decision-making and treatment plans \cite{Bader1998}. For example, self-logged data can "fill in the gap" between doctor visits \cite{West2016}. However, many barriers hinder a user (ordinary, patient, or doctor) from extracting maximum benefit from the wealth of data that can be acquired through self-logging. Some of these barriers include time constraints, poor visualization/analytics tools, limited analytical skills of the average user, and fragmented data across multiple platforms  \cite{Bentley2013}, \cite{Choe2014}, \cite{Li_CHI2010}. Wyatt and Wright \cite{Wyatt1998} highlight four key difficulties related to interpreting large amounts of health data, two of which are: 1) comparing within the dataset that is often spread across pages and sections, and 2) identifying trends. These barriers inform one of the objectives of this paper which is to design and evaluate an non-standard approach for visualizing, interpreting, and extracting clinically-relevant insights from daily records of CGM data in diabetes management.




\subsection{Data Visualization in Health}

One of the primary objectives of a visualization is to help people find valuable insights, where insights can be defined as "an individual discovery about the data by the participant, a unit of discovery" \cite{saraiya2005insight}. According to Choe et al., there are 8 types of visualization insights including detail, self-reflection, trend, comparison, correlation, data summary, distribution, and outlier. In practice, visual tools are often used and found to be effective as an alternative or supplement to text-only presentation of health data.  Pictorial representations can be more engaging and enable easier understanding of key details. In healthcare, data visualization has been used to communicate medication instructions \cite{Houts2006}, \cite{Katz2006}, health risk information \cite{Ancker2006}, treatment choices \cite{Hawley2008} and for just-in-time interventions \cite{Sharmin2015}. Earlier work by Powsner and Tufte's \cite{Powsner1994} presented a short but detailed graphical summary of patient data and status for quick access to the most relevant information in medical records. The approach did not aim to replace traditional methods of record-keeping, but to supplement it in ways that "allow the viewer to assess relations" and consider alternative management strategies when needed. Other notable visualization projects on which many variations have been built include \textit{LifeLines} \cite{Plaisant2003} and \textit{TimeLine} \cite{Bui2007}. For example, Aigner and Miksch \cite{Aigner2006} developed \textit{CareVis} by employing multiple views (i.e. logical, temporal, and a quickview panel) to organize and present patient data. More detailed reviews as it relates to knowledge discovery, information visualization, and challenges in medicine can be found in \cite{Chittaro2001}, \cite{Faisal2013}, and \cite{Holzinger2012}. From the wealth of knowledge in visualization, a key lesson for developing a useful tool is that end-users should be involved in the design process from conception \cite{Aigner2006}, \cite{Wyatt1998}.


\subsection{Decision Support in Diabetes Management}

\begin{figure*}
	\centering
	\includegraphics[width=0.7\textwidth]{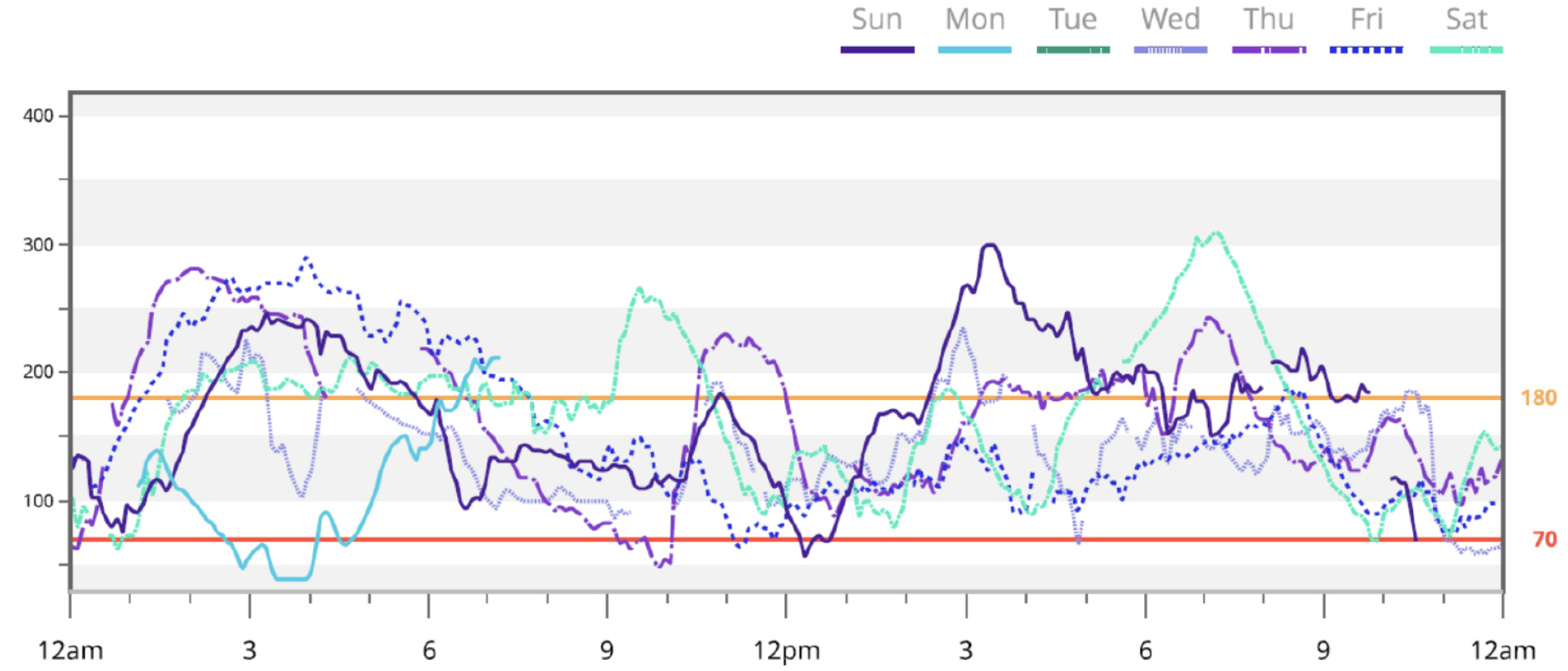}
	\caption{Example of the conventional daily overlay plot for data visualization and interpretation of CGM data  \cite{DexcomClarity2018}. }
	\label{fig:Dexcom_OverlayPlot}
	
\end{figure*}


Given that many daily decisions directly influence diabetes, there is great interest in methods that can ease the decision making process for all stakeholders. This includes tools to track and monitor behaviors that affect blood glucose, communicate management status, encourage reflection, and influence treatment plans. A review of mobile-based self-management applications can be found in the papers by Arsand et al. \cite{Arsand2010} and Tatara et al. \cite{Tatara2009}. Recent work by Katz et al. suggest that low adoption of diabetes management applications could be influenced by interfaces failing to address cognitive and emotional requirements of users \cite{katz2018data}. Specific to visualization of diabetes data, Frost and Smith \cite{Frost2003}, \cite{smith2007} introduced and evaluated a color-coded grid-format presentation, plotting time-of-day in the rows versus glucose measurements per day in the columns. Their tool enabled the user to "zoom in" to particular days to view a line-graph of glucose measurements with associated imagery, when present. Similarly, Desai et al. investigated various visual representations for forecasted post-meal blood glucose for persons with type 2 diabetes \cite{desai2018pictures} while Feller et al. \cite{Feller2018visual} introduced \textit{Glycolyzer} for analyzing the effect of meals on blood glucose measurements. \textit{Glycolyzer} was developed to contain a heatmap view of meals versus macronutrient and glycemic impact, meal images and a summary nutritional content, and probability density plots of glucose changes in response to nutritional content. This tool was designed for and evaluated by one stakeholder (i.e. registered dietitians) in a simulated clinical visit. Unlike many of the aforementioned studies, which rely on sparse blood glucose data from glucometers with about 4 - 8 measurements per day, our work uses data from CGMs with up to 288 samples per day.


\par A limited number of studies have developed tools for use with CGM data. Rodbard \cite{Rodbard2009} proposed using stacked bar charts to visualize glucose distribution in multiple ranges that facilitate comparison by date, time of day, and day of week. However, the usefulness of Rodbard's approach is unknown because their methods were not evaluated amongst end-users. In practice, a 24-hour (daily) glucose overlay plot is most commonly used to present CGM data \cite{Scheiner2016}. It is a 2-D time-series plot (BG versus time) included in most download reports by device manufacturers such as Dexcom \cite{DexcomClarity2018}, and Medtronic \cite{MedtronicCareLink}. As shown in Figure \ref{fig:Dexcom_OverlayPlot}, the daily overlay plot presents 7-days of data where each trend line represents a different day's blood glucose data. Additionally, horizontal lines that run parallel to the x-axis are used to indicate the target glucose range. Alternatively, experts within the diabetes community also use the Ambulatory Glucose Profile (AGP) \cite{Bergenstal2013}, \cite {Danne2017}. A single page report that includes "summary statistics, a glucose profile graph and an insulin profile graph or glucose daily calendar graphs" \cite{AGP}. Both the daily overlay plot and glucose profile plot in AGP reports are used to present up to 2-weeks of diabetes data for review and analysis. Unlike these visuals that graph raw glucose data in a trend or summary plot, in this paper, we evaluate PixelGrid, a color-based matrix-based plot with the aim of supporting end-users to identify temporal patterns.


\section{Method}

\subsection{Participants}
A total of 20 subjects (16 female, 4 male; ages: 19 - 45yrs) participated in this study approved by the appropriate Institutional Review Board (IRB). Subjects represented 3 end-user populations (i.e. stakeholders) including: patients with diabetes who use CGMs for daily management (n = 11), care-givers of patients with diabetes who use CGMs for daily management (n = 3), and clinicians of patients with diabetes who use CGMs for daily management (n = 6).  Of the clinicians who participated in this study, all but one are endocrinologists with diabetes speciality. Participants were recruited from a metropolitan area in Texas and online diabetes communities. 

\subsection{Dataset} \label{sec:Dataset}
We created all visuals for this study using one randomly-selected subject from a dataset including 60 days of CGM data from 10 persons with T1D. All CGMs recorded at a rate of 1 sample every 5 minutes. Given that CGMs are wearable systems for which the subject decides if/when to wear it, the amount of data samples per subject varies depending on the wear time per subject. The complete dataset includes 152,477 total CGM samples for 10 subjects. 


\subsection{Design of PixelGrid Visualization} \label{sec:PixelGrid_Design}

\subsubsection{CGM-Metrics for Visual Presentation} \label{sec:Metrics}
An effective visual analytics tool should enable quick access to key insights of relevance to the viewer. With this in mind, two sources informed the CGM metrics included in our visualization: 1) diabetes expert panel recommendations \cite{Bergenstal2013} \cite{Danne2017} \cite{Fonseca2016}, and 2) user input from our needs assessment survey (section \ref{sec:end-user_questionnaire}). We compiled and prioritized key metrics in the order listed below:
\begin{enumerate}
	\item \textit{Low Wear-Time} of CGMs is associated with decreased glycemic control \cite{Tamborlane2008}, \cite{vhaduri2020adherence}, \cite{Wong2014_DiabCare}. According to Rodbard \cite{Rodbard2016}, "benefit is proportional to frequency of use." Therefore, low wear-time, defined in this work as less than 50\%, ranked as the first priority metric to track and display in our visualization. Daily glycemic control cannot be quantified when there is no supporting data of a patient's blood glucose trends. 
	\item \textit{Low Blood Glucose} is directly associated with acute complications such as loss of consciousness, seizures, and even death \cite{ADA2017}, \cite{Bergenstal2013}, \cite{faiola2018hypoalert}. Following analysis for data sufficiency, identifying hypoglycemic (i.e. low blood glucose) events is a priority. Diabetes experts recommend gradation of low blood glucose into buckets of \textit{low} (54 - 70 mg/dL) and \textit{very low} (< 54 mg/dL) \cite{Bergenstal2013}, \cite{Danne2017}. Therefore, these two metrics were included in our visualization for quick access by the end-user, with higher priority given to \textit{very low} blood glucose because its consequence is more severe.
	\item \textit{High Blood Glucose} is associated with long-term vascular complications \cite{ADA2017}. In addition to assessing hypoglycemic events, decreasing the occurrence and severity of hyperglycemic (i.e. high blood glucose) events is a tertiary concern. Similarly, diabetes experts recommend gradation of high blood glucose into buckets of \textit{high} (> 180 mg/dL) and \textit{very high} (> 250 mg/dL) \cite{Bergenstal2013}, \cite{Danne2017}. Therefore, these two metrics were included in our visualization for quick access by the end-user, with higher priority given to \textit{very high} blood glucose because it is more severe.
	\item \textit{Time-in-Range} is an important metric for quantifying glycemic control in diabetes management. It refers to the percentage of reading in the target range of 70 - 180 mg/dL \cite{Bergenstal2013} \cite{Danne2017}. Following analysis of the above suboptimal events, time-in-range was calculated and categorized as positive when greater than 70\%. 
\end{enumerate}

\subsubsection{Layout of PixelGrid Visualization}

\begin{figure}
    \begin{subfigure}{0.5\textwidth}
    \centering
    	\includegraphics[width=0.8\textwidth]{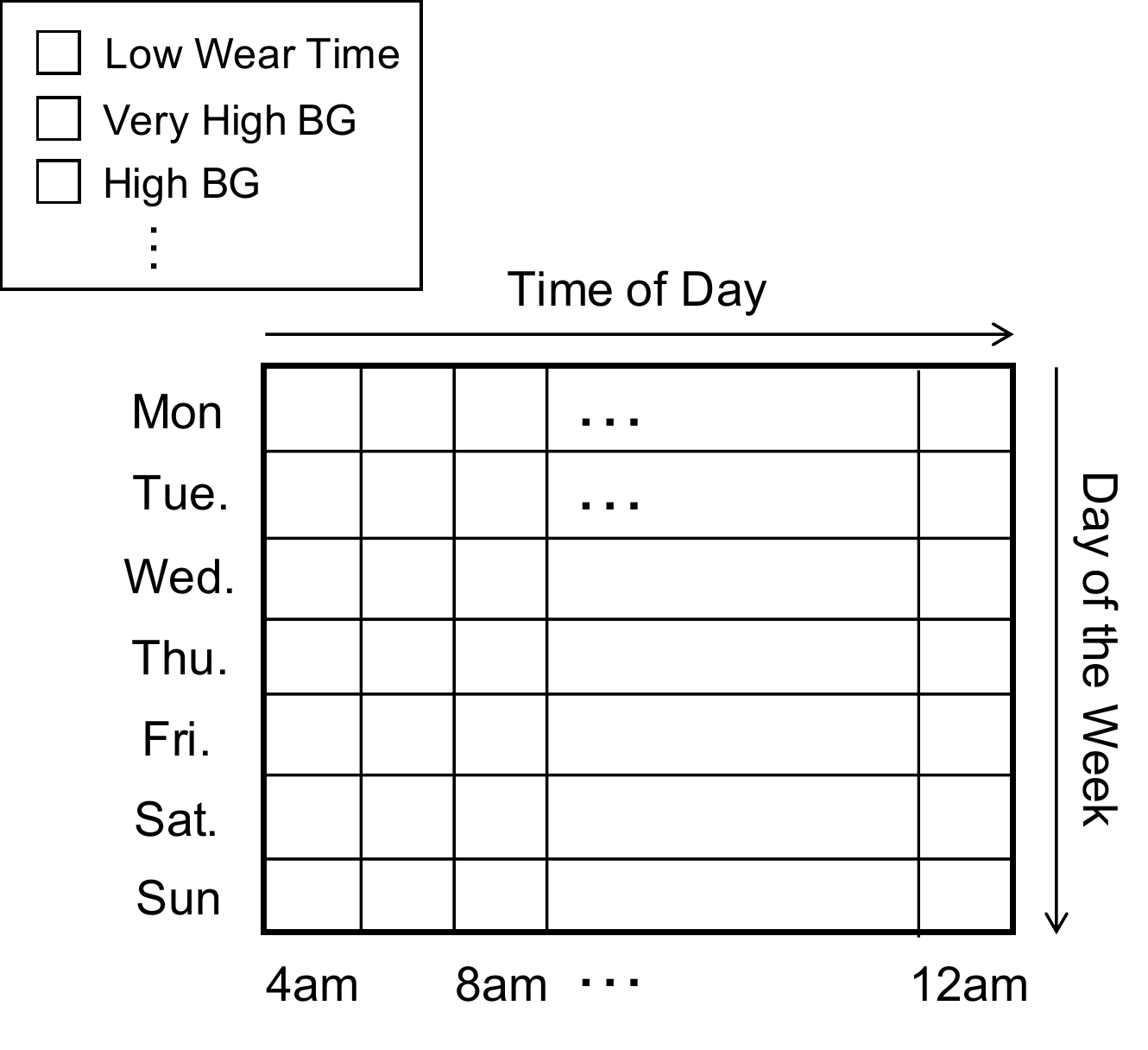}
    	\caption{Single-week assessment. BG means "Blood Glucose."}
    \end{subfigure}
    \par\bigskip
    \begin{subfigure}{0.5\textwidth}
    \centering
    	\includegraphics[width=0.8\textwidth]{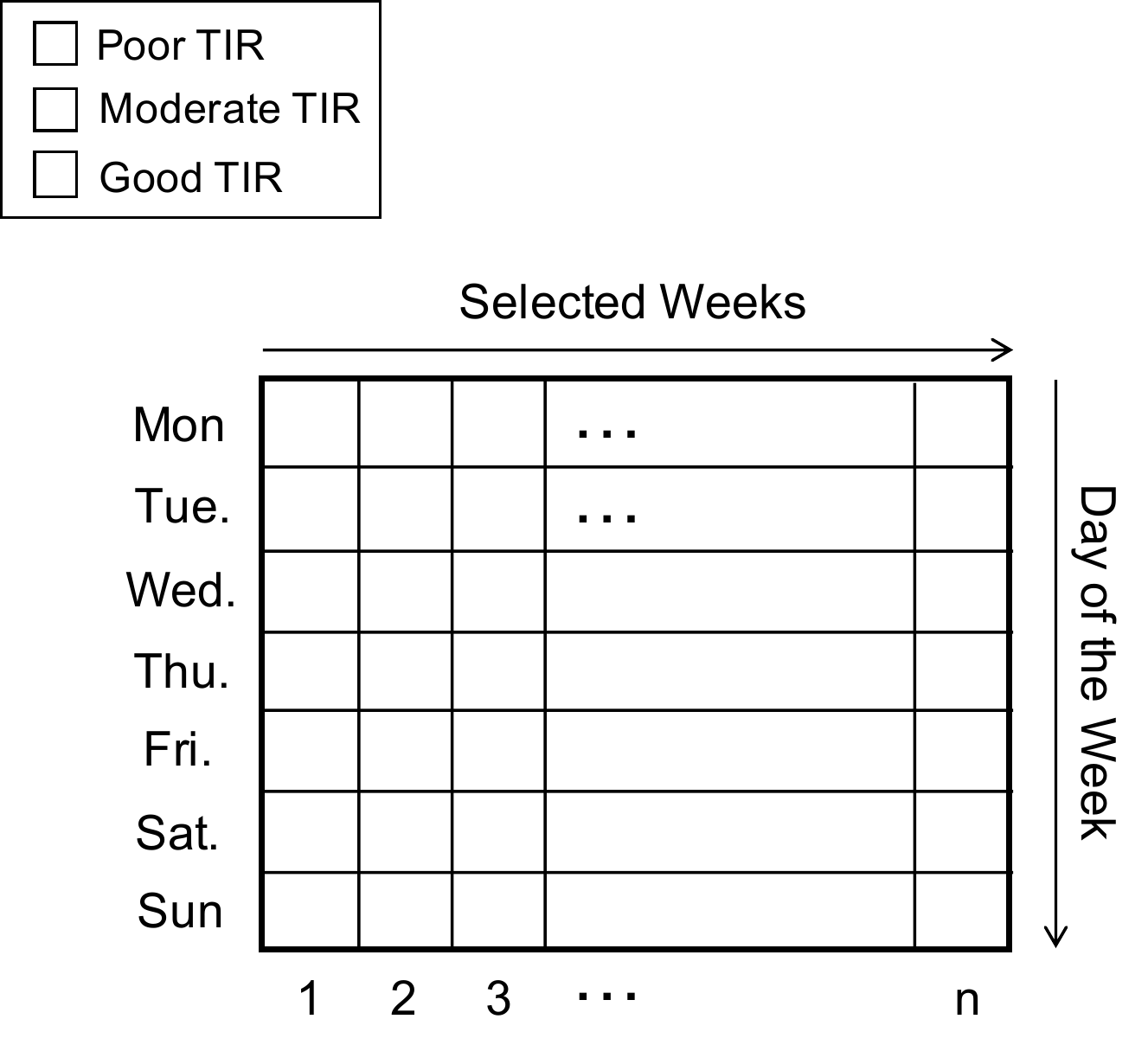}
    	 \caption{Multi-week assessment. TIR means "Time-in-Range."}
    \end{subfigure}
    \caption{Layout of PixelGrid visualization of CGM data.}
    \label{fig:PixelGrid_Sample}
\end{figure}

PixelGrid visualization presents key parameters for evaluating diabetes management in an $m \times n$ matrix-based plot. Prior research highlights the need for methods of displaying blood glucose data to facilitate comparison by date, time of day, and day of the week \cite{Rodbard2009}. Therefore, the chosen layout of PixelGrid presents CGM data under the confines of recurring time-blocks for quick-access pattern recognition. The layout chosen aims to enable rapid identification of habits and daily schedules that influence good and/or suboptimal glucose trends. 

\par We developed and evaluated PixelGrid plots for short-term (i.e. single week) and long-term (i.e. multi-week) assessment of real CGM datasets. In single-week visualizations, the rows were used to represent days of the week ranging from Monday to Sunday, while the columns were used to represent hours of the day ranging from 0 to 24 (i.e. midnight-to-midnight). Row assignments were kept consistent in the multi-week visualization to minimize cognitive load. However, the columns of multi-week plots were used to represent user selected weeks for analysis. Figure \ref{fig:PixelGrid_Sample} shows a representative image of the final (Phase 2) layout of PixelGrid visualization for single- and multi-week CGM assessment.

\par Contrasting colors were also used to represent each discrete parameter that could be of interest to an end-user retrospectively reviewing a CGM dataset. Prior research \cite{Healey1996} identified 3 color-selection criteria that enable an observer to rapidly and accurately search a visual, namely, color distance, linear separation, and color category. Additionally, a comparison study of the trade-off between the number of colors in a display and response time to identify a target found that the human visual system can handle up to five different colors in one image \cite{Healey1996}. These results informed the choice of colors in PixelGrid visualization. The neutral color, white, was used to represent periods where the CGM contained insufficient data points for analysis, this is indicative of a user not wearing their device (i.e. low wear time). Warm colors, namely orange and yellow, were used to represent \textit{very high} and \textit{high} blood glucose, while, cool colors, namely, dark blue and light blue were used to represent \textit{very low} and \textit{low} blood glucose. The color green was used to represent periods where the user's blood glucose was maintained within the target range (i.e. good diabetes management). Unlike the work by Desai et al. \cite{desai2018pictures}, we avoided using the color red for very high or very low values as this can elicit fear in users and end up being more discouraging than encouraging.

\section{User Study and Results}
\begin{figure*}
    \centering
	\includegraphics[width=0.95\textwidth]{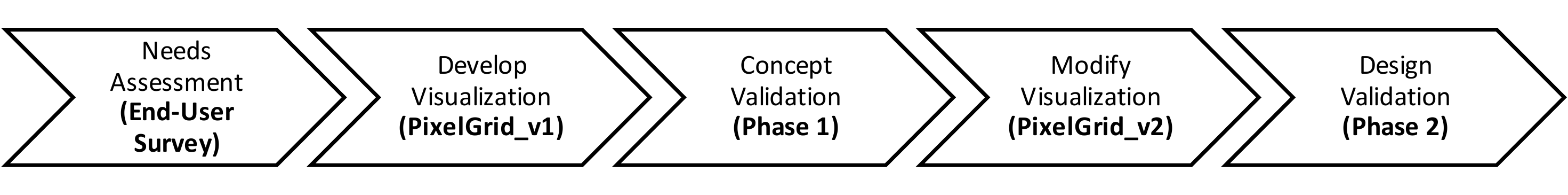}
	\caption{Overview of PixelGrid Design Process}
	\label{fig:DesignOverview_PixelGridViz}
\end{figure*}

Figure \ref{fig:DesignOverview_PixelGridViz} shows an overview of the multi-stage design process. All participants started by completing an end-user survey tailored for patient/care-giver versus clinician population. This survey was to understand needs within a community of stakeholders and for inform choices in developing PixelGrid visualization. Five subjects (3 patients with diabetes and 2 endocrinologists with diabetes specialty) participated in the concept validation (i.e. Phase 1) user study. Feedback collected from Phase 1 was used to inform modifications toward developing an improved visualization that was evaluated in a second user study. Phase 2 was fully completed by 14 subjects (7 patients with diabetes, 3 care-givers, and 4 clinicians with diabetes specialty). One subject provided incomplete data resulting in 14 instead of 15 subjects in Phase 2. There was no overlap between subjects who participated in Phase 1 and Phase 2 user study. In addition, participants received \textit{no training} and \textit{had no exposure} to PixelGrid visualization prior to participating in this study. 

\subsection{End-User Needs Assessment Survey} \label{sec:end-user_questionnaire}
This survey contained two open-ended questions crafted to understand: 1) what users focus on when they review retrospective glucose data? and 2) what users view as existing gaps in state-of-the-art diabetes data reporting tools? All participants were familiar with CGM reports and had extensive experience (1 to > 10 years) with daily overlay plots (see Figure \ref{fig:Dexcom_OverlayPlot}) for reviewing and reflecting on diabetes-device data. Given this, the below questions were asked:

\subsubsection{\textbf{When reviewing a CGM report, what are the top things you look for in recent management?}}
The top 5 responses that emerged from participants include: time-in-range (9 mentions), patterns/trends of highs and lows (9 mentions), average blood glucose (6 mentions), trends during certain times/hours/days (5 mentions), and time-below-range (4 mentions). These responses are consistent with key metrics identified by diabetes experts in consensus reports \cite{Bergenstal2013} \cite{Danne2017} \cite{Fonseca2016}. Subject 5 (physician) stated that they focus on: \textit{1) time-in-range, 2) percent in hypoglycemia (< 54mg/dL), and 3) daily trends}. Subject 6 (patient) stated that they look for \textit{trends by time of day, average blood glucose, and trends of high/low activity/stress days}. Other prevalent responses that subjects mentioned include: effect of activity such as exercise/meal/insulin, glucose spikes/variability and standard deviation. The visualization presented in this paper does not encompass all responses, however, it includes majority of the most prominent responses. 


\subsubsection{\textbf{What do you wish CGM reports included that could help and/or speed up analysis and interpretation of diabetes management?}}
Two prominent responses emerged from participants including a desire for CGM reports to: 1) automatically extract trends/patterns, detect changes in management, and suggest specific behavior changes (6 mentions), and 2) integrate data from insulin pumps, bluetooth enabled insulin pens, and carbohydrate intake (6 mentions). Subject 14 (patient) stated that they wish CGM reports identified \textit{any recent changes in management (for example: time-in-range has decreased in [the] last week}. On the other hand, subject 15 (patient) stated \textit{it would be helpful if it [CGM reports] were more clear identifying when/where I should be making specific changes to basal or bolus [insulin] choices}. Other prevalent responses mentioned include: the ability to review more days in the same graph, identify best/worst days of management to understand what behaviors to avoid, and not having to upload supplemental data or manually enter data into reports. Subject 12 (patient) said \textit{I wish it [CGM reports] integrated data from my [insulin] pump without me having to manually enter it... It would also be cool to know what foods I ate etc. without manually entering it. It is too tedious to manually enter it in, therefore I never do it.} The visualization presented in this paper does not encompass all responses, however, it is a building block toward 3 desires mentioned, namely, automatically extracting trends/patterns, enabling review of multiple days in the same graph, and identifying the best and worst days of management to pin-point supporting behaviors. 

\subsection{Insights for Designing Decision Support Tools}
Themes identified from our needs assessment survey suggest three key things that users desire to support meaningful reflection on personal health data, namely:

\begin{enumerate}
    \item \textit{Specific Summarization Metrics}: These metrics are often familiar quantities that are easy to calculate such as time-in-range (i.e. a ratio) and average blood glucose (i.e. mean of a set of numbers). Additionally, these metrics should be relevant to quantify "history" and "status" \cite{Li2011_ubicomp} such as patterns/trends of highs and lows, and trends during certain times/hours/days.
    \item \textit{Solutions that Minimize Cognitive Effort}: This is evident from users desire for data reporting tools to "automatically" extract trends/patterns, detect change in management, and suggest specific behavior change. The point here is automatic summary of status, identification of recent changes, and recommendation for improvement. 
    \item \textit{Solutions that Enable Data Integration}: It is not uncommon for people to use multiple self-tracking devices for their unique capabilities in today's society. Therefore, users desire for solutions that enable integration of personal health data for multiple sources to support a greater understanding of interrelated metrics. For example, physical activity has a direct effect on blood glucose, hence, solutions that support integration of these data sources would be useful.
\end{enumerate}

\section{Evaluation of PixelGrid}
The following section reports on a two-phase evaluation of the proposed PixelGrid visualization described above and insights identified for designing effective data visualization tools for personal health data.

\subsection{Phase 1: PixelGrid Concept Validation} \label{sec:phase1}
\begin{figure}
	\includegraphics[width=0.5\textwidth]{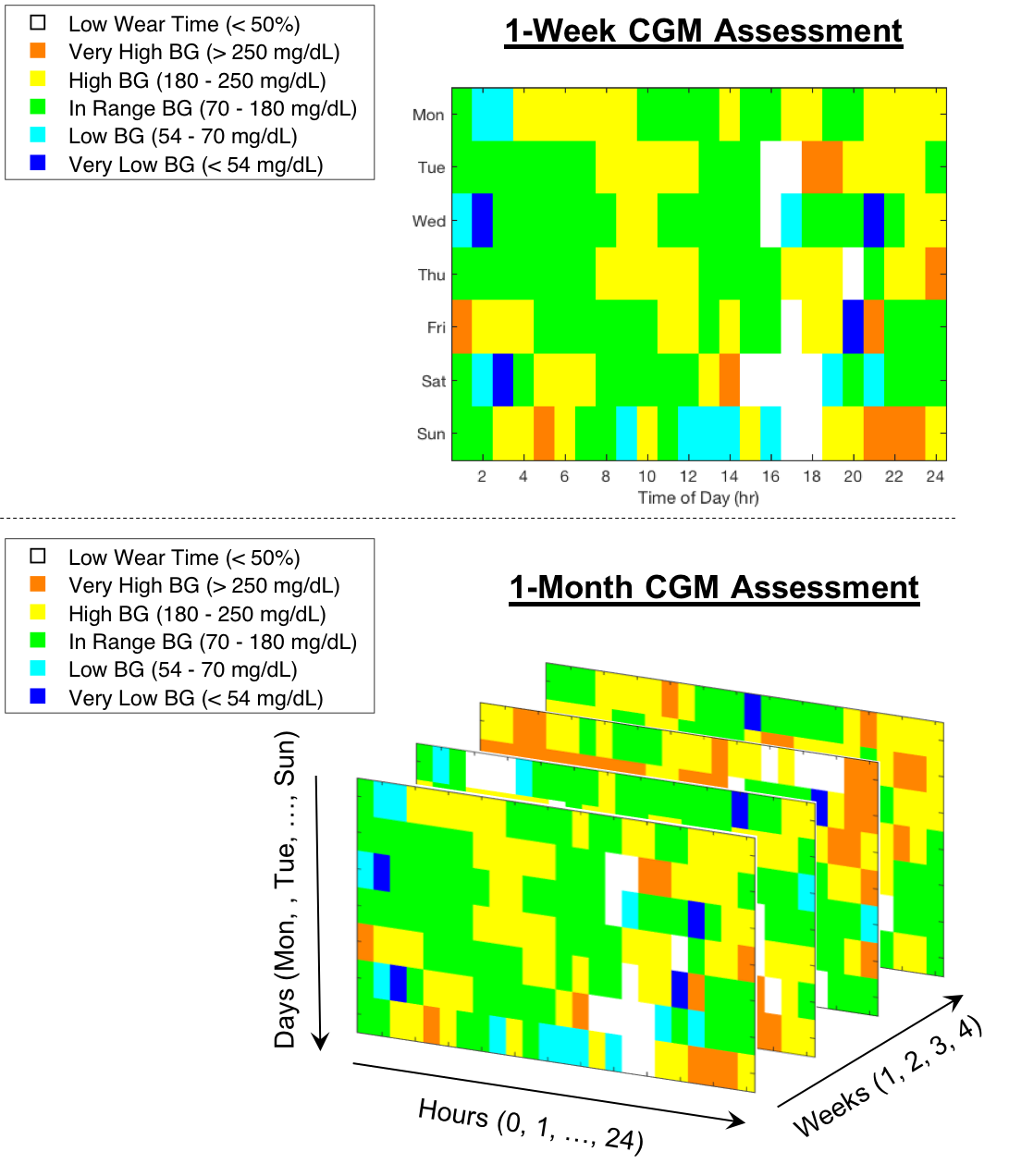}
	\caption{Phase 1 concept validation: PixelGrid visualization of real CGM dataset}
	\label{fig:Phase1_PixelGridViz}
\end{figure}

A remote design platform, UsabilityHub \cite{UsabilityHub}, was used to conduct this concept validation study. Real blood glucose data as described in section \ref{sec:Dataset} was presented using PixelGrid visualization. Figure \ref{fig:Phase1_PixelGridViz} shows the two PixelGrid images created to depict 1-week (i.e. short-term) and 1-month (i.e. long-term) CGM data for retrospective analysis. Subjects reviewed this image while answering 4 open-ended questions.
During this phase, 3 participants were observed remotely and asked to "think aloud" \cite{Charters2003} while a researcher took notes of pain points not captured in users' textual response.

Below are the questions and learnings from Phase 1 testing:

\subsubsection{\textbf{Looking at the proposed visualization, what do you understand from it?}} The majority of the participants, 4 out of 5 (or 80\%),  identified salient details of blood glucose readings as it relates to times/days when it was most prominent. Subject 2  (patient) stated, \textit{I can see a pattern of highs from 8am to 10am ... there are also small patterns of low [blood glucose] after highs when highs occur after 3pm}. Subjects 1 (physician) and 4 (patient) noticed that \textit{low BG episodes mostly occur on Sundays, mid morning to late afternoon}. This connection between key metrics of diabetes management and time of day/week can support identification of temporal patterns that negatively or positively influence blood glucose trends. According to subject 3 (patient), \textit{this visualization lets you see where there are clusters of highs and lows and whether they remain fairly consistent over the course of weeks or months}. 

\subsubsection{\textbf{What is a scenario where the proposed visualization might be important?}} All participants, 5 out of 5 (or 100\%), mentioned that the PixelGrid visualization is useful for quick identification of trends and recurrences that happen during certain days or hours. As a physician, subject 1 stated that this visualization \textit{allows the reviewer to ask specific and directed questions to the patient about certain dates or experiences that resulted in glycemic excursions}. Subject 5 (physician) stated, this visualization can \textit {help guide diabetes treatment regimens (insulin doses, exercise, etc).} Subject 3 (patient) stated that \textit{this visual might be useful if you want to view a lot of data over a long period of time but do not want to condense it all into one trend and end up missing trends throughout the week... The color blocks are simple and straightforward.}

\subsubsection{\textbf{What do you like about the proposed visualization tool?}} Subjects liked that this visualization is: 1) color-based, 2) enables the user to quickly find patterns and compare data from different times/days, and 3) does not overlay information on top of each other (i.e. not cluttered). Subjects 2, 3, and 5, all used the word \textit{easy} in their description of what they like. Subject 2 (patient) stated that PixelGrid visualization is \textit{a less-crowded method of seeing multiple days and times in one plot for easy comparison}. Subject 3 (patient) stated that \textit{making the data look like blocks makes it easy to look at where high and low [blood glucose periods] are clustered. I appreciate how easy it is to compare between weeks.} Subject 4 (patient) stated a similar point, \textit{that it is based on color instead of numbers. Graphs with dips and highs can make it confusing when you're overlaying information.} 

\subsubsection{\textbf{What additional features/changes would you like to see added to the proposed visualization?}}Recommendations mentioned by more than one subject include: 1) adding summary metrics of patterns in a text-based format to the visual, and 2) enabling an interactive data visualization tool such that an end-user can select the dataset to plot and ranges for high/low blood glucose. Subject 1 (physician) suggested to add \textit{percent of time for each category written in prose alongside the image}. Similarly subject 5 (physician) suggested to \textit{summarize patterns}. The other key point of an interactive data visualization is supported by subjects 2 and 3. Subject 2 (patient) stated that \textit{it might be useful to do [select] a day (i.e. a month of all Mondays)} and subject 3 (patient) stated \textit{maybe a slider to decrease the range for each group would be helpful so people who want more information can get it.}

\subsection{Phase 2: Improved PixelGrid Validation} \label{sec:phase2}
\begin{figure*}
    \centering
	\includegraphics[width=0.8\textwidth]{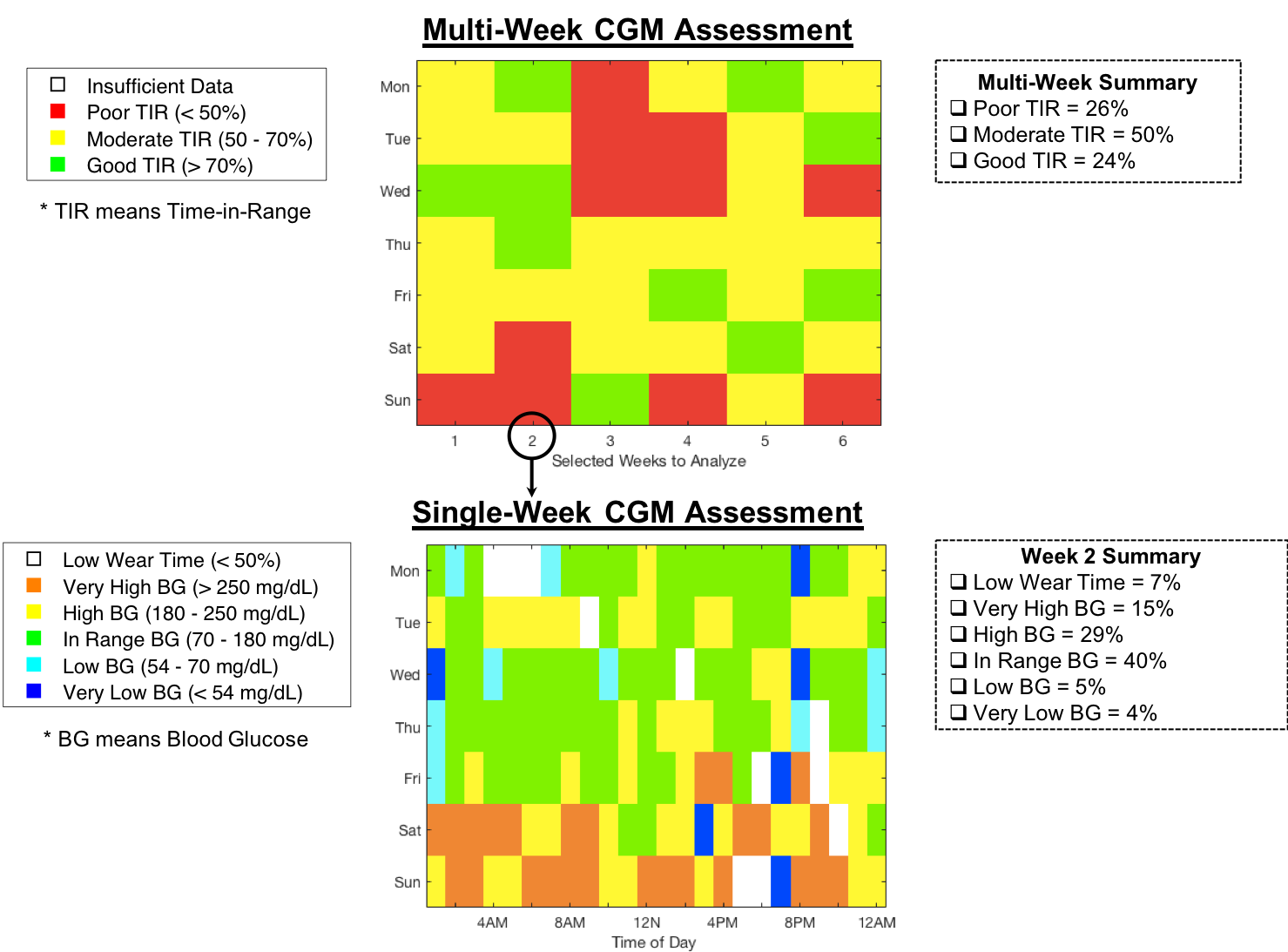}
	\caption{Phase 2 concept validation: PixelGrid visualization of real CGM dataset}
	\label{fig:Phase2_PixelGridViz}
\end{figure*}

Figure \ref{fig:Phase2_PixelGridViz} shows an image of the improved PixelGrid visualization that was developed and evaluated. In this phase, a new set of 15 subjects were asked open-ended questions about the design. UsabilityHub \cite{UsabilityHub} was used for remote testing. This phase did not include remote observation and the "think aloud" method. Based on the feedback received from Phase 1, 4 key items were  modified to create the Phase 2 visualization. 

\begin{description} 
	\item[Modification 1] Using a single image for long timescale (multi-week CGM) visualization. This revision was made to reduce the cognitive load required by an end-user to find patterns from multiple images. From phase 1 testing, 3 participants, subjects 1, 2 \& 4, mentioned that the 1-month CGM assessment (bottom image in Figure \ref{fig:Phase1_PixelGridViz}) was hard to follow. Subject 1 (physician) stated that it \textit{is challenging to see all the data in the current image [1-month CGM assessment]}.
	
	\item[Modification 2] Adding summary metrics in a text-based format to the visual. This feedback was directly mentioned by 2 participants in Phase 1 testing. Subject 5 (physician) stated in response to suggested changes: \textit{summarize patterns, for example, time of day that CGM user is running high/low.} Similar to modification 1, it is expected that this change will reduce the cognitive load required by an end-user to mentally calculate summary metrics for quick understanding of management.
	
	\item[Modification 3] Enabling the end-user's control of what data to view in more details. This revision allows the user to select a specific week to further analyze from the multi-week CGM assessment plot. From Phase 1 testing, subject 3 (patient) suggested a change to accommodate for \textit{people who want more information}.
	\item[Modification 4] Changing x-axis labels from a 24-hour time convention (less common in the U.S.) to a 12-hour time convention (more common in the U.S.). In Phase 1 testing, subjects 1 (physician) \& 2 (patient) identified patterns from Figure \ref{fig:Phase1_PixelGridViz} in specific time ranges, however, they used A.M. and P.M convention to report their observations. This means the participants likely expended unnecessary efforts to convert from the 24-hour clock to the 12-hour clock. This change appeals to the cultural norm and context of users.
	
\end{description}

Specific questions and learnings from Phase 2 are as follows:

\label{sec:open-ended_PatternQ} \subsubsection{\textbf{Take a minute to study the proposed PixelGrid visualization, what insights can you draw about the user's diabetes management?}} The majority of participants, 12 out of 14 (or 85.7\%), consistently identified the most prominent insight, which is that the user has poor management on weekends in the single-week plot compared to weekdays. Additionally, 7 out of 14 (or 50\%) participants consistently identified a second insight that the user has a pattern of low BG in the evenings. These are both clinically-relevant patterns useful to inform future treatment plans. Given the relevance of these patterns in the underlying data, two participants recommended a change in treatment regimen during certain time periods. For example, subject 8 (patient) stated, \textit{from the single-week [CGM assessment plot], it looks like perhaps weekend [insulin] rates/management could be adjusted... look at changing overnight basal rates to avoid sleeping low [i.e. low BG during the night]... consider raising basal [insulin] rates in the evening as well, as there seems to be a pattern for very low BG.} A similar comment was made by subject 18 (physician) who stated that the patient \textit{has fair control during the week, however poor control on weekends. [There are] patterns of lows in evening, so need to ask if patient is exercising, how much [they are] eating, etc. to ascertain the reason.} Such responses align with the objective of solutions to support reflection on personal health data. These responses support that PixelGrid enables end-users to identify patterns of good and/or poor management from retrospective CGM data. Additionally, some participants found PixelGrid useful to guide questioning to identify behaviors associated with management gaps, and to support treatment regimen changes that can address future mishaps.

\label{sec:ListPatterns} \subsubsection{\textbf{List all of the patterns you see in: [1] PixelGrid "Multi-week CGM Assessment" (top) plot, and [2] "Single-week CGM Assessment" (bottom) plot. Leave blank if none.}} A total of 31 patterns were identified by study participants ranging from a minimum of 0 (i.e. blank response) to a maximum of 7 patterns. At least 3 patterns were identified by 7 out of 14 (or 50\%) participants. Only 2 subjects (or 14\%) left this response blank (i.e. identified 0 patterns). From the multi-week plot, subject 12 (physician) stated that the user had \textit{poor TIR on Sundays, pretty good [TIR] on Thursdays and Fridays. [The user's BG was] most variable on Tuesday and Wednesdays. Week 5 was the best week.} Additionally, from the single-week CGM plot, subject 12 (physician) observed \textit{good [CGM] wear time (93\% of the time), higher BG on Saturday and Sunday. More lows [i.e. low BG] toward the evening.} These responses also support that PixelGrid was useful to most participants for identifying good and suboptimal patterns in diabetes management.

\label{sec:PixelGridDislikes} \subsubsection{\textbf{What do you like or NOT like about the PixelGrid visualization?}} There was a consensus on two key points relating to what participants did not like about PixelGrid. First, it provides no specific recommendations about what the user can change to improve their diabetes management. Subject 19 (patient) stated that they \textit{hated the multi-week [CGM assessment plot] because it's hard to understand what to change}. Similarly, subject 20 (physician) stated that PixelGrid \textit{gives a general pattern but no specifics as to what needs to be changed.} The second area for improvement identified by participants relates to the choice of colors in PixelGrid visualization. Subject 12 [physician] stated \textit{if there are no detectable patterns, you [the end-user] will just see random colors which could be frustrating to a patient [end-user].} Likewise, subject 17 [physician] stated \textit{I think for some, the color flow may be confusing or not intuitive.} Another response for consideration in future work mentioned by subject 16 (physician) is to add more data/details on hypoglycemia (low BG). On a positive note, subject 18 [physician] stated \textit{I do like how I can visualize overall patterns in ONE screenshot.} The above feedback will be used to inform future work on designing effective decision support tools for diabetes management.

\section{Insights for Designing Data Visualization Tools}
Themes identified through the design and evaluation of PixelGrid suggest four key things that users desire in data visualization tools to support meaningful reflection on personal health data, namely:

\begin{enumerate}
    \item \textit{Complementary Textual Descriptors}: Graphic presentations alone are not sufficient to support reflection on personal health data which can be large and multidimensional. Therefore, effective data visualization tools should include complementary textual descriptors (e.g. summary metrics) of the visual content. From the concept validation phase of our design and evaluation process, multiple subjects recommended adding summary metrics in a text-based format to the graphical presentation to communicate key takeaways.
    \item \textit{Concise and Flexible}: Users desired to have a minimum number of visuals and content in a single-view. However, they also desired flexibility with regards to what content can be viewed. For example, in the first phase of our study, subjects 2 and 3 alluded to enabling user control "so people who want more information can get it." Such flexibility can be provided with an interactive as opposed to static solution. 
    \item \textit{Contextually-Fitting}: The format for data reporting should be tailored to fit the contextual norm of user populations. For example, users in this study preferred to see the timescale reported using a 12-hour clock (i.e. A.M. and P.M.) instead of a 24-hour clock, however this preference could be reversed for users in a different culture or setting where a 24-hour clock is the standard. 
    \item \textit{Informative and Actionable}: Effective data visualization tools should be both informative and actionable as one without the other provides little value to users. In our study, users that identified actionable insights from information presented using PixelGrid rated the visualization more positively while users who found PixelGrid to be informative but not actionable rated the visualization more negatively.
\end{enumerate}

\section{Summary and Discussion}

In this paper, we strive to understand reflection needs of users of wearable medical devices, one of which is a continuous glucose monitor used in diabetes management. To learn insights of real end-users, we included patients with T1D, care-givers, and clinicians in our multi-stage study. From our needs assessment survey, we identified 3 important things that users desire from decision support tools to support meaningful reflection on personal health data. Firstly, users desired to have \textit{specific summarization metrics} that are familiar quantities and easy to understand. Specific to diabetes management the top 5 evaluation metrics desired were time-in-range, patterns/trends of highs and lows, average blood glucose, trends during certain times/hours/days, and time-below-range. Additionally, users desired \textit{solutions that minimize cognitive effort} and \textit{solutions that enable data integration}, both of which were evident at various points in our study. For example, the most prominent items that emerged from user's wish lists are solutions that automatically extract trend/patterns, detect changes in management, and suggest behavior changes. The other wish list item that ranked high was for solutions that can integrate pertinent data from multiple devices. These insights are critical to consider and incorporate in the design of effective design support tools that facilitate reflection on previous health to inform and improve outcomes. 


\par Following this, we developed and evaluated a data visualization tool with the objective of enabling users to quickly and easily identify temporal patterns in diabetes management. Our proposed visualization takes a non-conventional approach compared to what is used in practice (i.e. daily overlay plots shown in Figure \ref{fig:Dexcom_OverlayPlot}). The PixelGrid approach presents blood glucose data in the various categories of interest (e.g. high, normal, low, etc.) using a color-based matrix plot. From the evaluation, we found that majority of users (84\% across 2 studies) found it useful for identifying salient patterns as it relates to times/days in blood glucose data. A few subjects also found PixelGrid useful to pin-point habits that contribute to management gaps, and to inform positive modifications to treatment regimen. For example, subject 18 (physician) stated some questions they would ask to identify factors influencing a management gap when they mentioned: \textit{there are patterns of lows in the evening, so need to ask if patient is exercising, how much [they are] eating, etc. to ascertain the reason.} Meanwhile, subject 8 (patient) recommended 2 changes to treatment regimen by stating: \textit{from the single week CGM assessment plot... look at changing overnight basal rates to avoid sleeping low [i.e. low BG during the night]... consider raising basal [insulin] rates in the evening as well, as there seems to be a pattern for very low BG}. These results are in alignment with the primary objective of developing solutions that support reflection to enable users learn from past data to inform future decisions.


We found that most participants liked the color-coded matrix-based layout for visualization. For example, subject 2 (patient) stated that PixelGrid \textit{is a less-crowded method of seeing multiple days and times in one plot for easy comparison}, while subject 3 (patient) stated \textit{making the data look like blocks makes it easy to look at where high and low [blood glucose periods] are clustered.} However, subject 12 (physician) identified a potential drawback of this presentation style which is that when there are no prominent patterns/clusters, the visual can become busy with many seemingly random colors. Further research is needed to identify an appropriate solution to this problem by carefully selecting the color choices and expanding the research to include automatic pattern recognition to guide users to relevant insights. A key take-away is that participants appreciated a summary visual of long-term data in single view, however they also wanted explicit text-based summary of the content as was learned from feedback received in the Phase 1 user study.

Through our two-phase design and evaluation process, we identified four important things that users desire in data visualization tools to support meaningful reflection on personal health data. Our findings show that users desire \textit{complementary textual descriptors} to support visual content, \textit{concise and flexible} presentation that support user control of what is in view, \textit{contextually-fitting} descriptors that match what is familiar, and \textit{informative and actionable} insights that identify what to change. In today's society, where personal health data is consistently growing, there is a critical need for understanding the needs of users and designing technology that can meet those needs. Wearable medical devices are becoming prominent for continuous monitoring of other health conditions. For example, wearable electrocardiograms such as iRhythm's Zio XT \cite{iRhythmZioXT}, are useful for continuous monitoring of arrhythmia. In the aforementioned example and beyond, solutions that support users with reflecting on personal health data will play a critical role in maximizing the benefit of these information sources. 


\subsubsection{Limitations}
This work has a number of limitations that should be addressed in future studies. One limitation is that in this work we did not show subjects a comparison using an existing visualization method. This choice was made because majority of our subjects (18 out of 20) were experienced users with 1 to more than 10 years of experience using the conventional daily overlay plot for interpretation of CGM data. Meanwhile, all users received no training and had no prior exposure to the presented visual tool used in this work. We believe the newness factor would have affected a fair comparison. However, subjects inherently compared PixelGrid to what they were familiar with and this is evident through their responses presented above. 

\par A second limitation is that based on the current design, PixelGrid presents data from a singular outcome variable (i.e. blood glucose) for retrospective analysis. However, it is known that users would find increased benefit in a tool that analyzes and presents insights from individual and related parameters including medication/insulin, food intake, activity, and even sleep. Brown \cite{Brown_42FactorsBG} presents a more comprehensive list of factors that affect blood glucose, many of which should be considered for monitoring and reporting to inform decision making in diabetes management. Responses from our end-user needs survey support the direction of integrating data from multiple sources. For example, subject 12 (patient) said \textit{I wish it [CGM reports] integrated data from my [insulin] pump... It would also be cool to know what food I ate etc. without manually entering it.} 

\par Additionally, this work used a static visual platform to conduct user studies. Previous research \cite{Bui2007}, \cite{Wyatt1998} and learnings from the Phase 1 concept validation study suggests that end-users desire an interactive system which allows for data manipulation and control of the content/level of detail presented. Our attempt to show an interactive visual (e.g. the user selecting week 2 for review in Fig. \ref{fig:Phase2_PixelGridViz}) on a static platform was missed by most participants. Furthermore, we did not enforce the use of any given platform during participation in our study, hence it is unclear how the platform used affected a user's experience. We observed that some subjects completed the study on their mobile device which was not ideal for looking at the visual while responding to related question. Succeeding studies will standardize the platforms used to eliminate this inconsistency.

\section{Future Work}

This paper is a launchpad for further research on decision support tools for diabetes management. Our developed solution was effective with enabling participants easily identify recurrent patterns present in the underlying blood glucose data. However, there is still room for improvement. Given the learnings from this study and desires of stakeholders (patients, care-givers, and clinicians), future work includes: 
\begin{enumerate}
	\item Automatic detection of bio-behavioral patterns in diabetes data to identify connections between blood glucose outcomes and the context in which they occur. This will in turn reduce the cognitive load required by end-users to manually identify patterns from large and continuously growing datasets.
	\item Evaluate the feasibility and accuracy of a recommendation system that provides specific suggestions based on retrospective data and patterns regarding what user can change to improve their diabetes management. 
\end{enumerate}

\balance{}

\bibliographystyle{SIGCHI-Reference-Format}
\bibliography{new_library}

\end{document}